\documentclass[fleqn,usenatbib]{mnras}

\usepackage{newtxtext}
\usepackage{newtxmath}

\usepackage{natbib}
\usepackage[T1]{fontenc}
\usepackage{aecompl}

\usepackage{graphicx}							
\usepackage{bm}									
\graphicspath{{./graphics/}}					
\usepackage{amsmath}							
\usepackage{amssymb}							
\usepackage{physics}							
\usepackage[nameinlink, capitalize]{cleveref}	
\usepackage{hyperref}							
\usepackage{array}								
\usepackage{xcolor}								

\newcommand{\erad}{\theta_\mathrm{E}}

\newcommand{\prob}[2]{\mathrm{P}\left(#1\vert#2\right)}

\newcommand{\Msun}{M_\odot}

\newcommand{\Mhm}{M_\mathrm{hm}}
\newcommand{\fsub}{f_\mathrm{sub}}
\newcommand{\Menc}{M_{2\erad}}

\newcommand{\musub}{\mu_\mathrm{sub}}
\newcommand{\mmin}{m_0}
\newcommand{\mmax}{m_1}
\newcommand{\mmap}{m_{\mathrm{map}_i}}
\newcommand{\Mtot}{M_\mathrm{tot,sub}^\mathrm{CDM}}

\newcommand{\vmax}{v_\mathrm{max}}
\newcommand{\rmax}{r_\mathrm{max}}
\newcommand{\Mmax}{M_\mathrm{max}}

\newcommand{\modelone}{\mathcal{M}_1}
\newcommand{\modeltwo}{\mathcal{M}_2}
\newcommand{\modelthree}{\mathcal{M}_3}
\newcommand{\mstrength}{\eta_m}
\newcommand{\mangle}{\phi_m}

\defcitealias{ORiordan2023}{O'R23}
\defcitealias{Despali2022}{D22}

\title[Angular complexity in substructure detection]{Angular complexity in strong lens substructure detection}

\author[C. M. O'Riordan et al.]{
	Conor M. O'Riordan$^{1}$\thanks{E-mail: conor@mpa-garching.mpg.de}, Simona Vegetti$^{1}$
	\\
	$^{1}$Max Planck Institut f\"{u}r Astrophysik, Karl-Schwarzschild-Stra{\ss}e 1, 85748 Garching bei M{\"u}nchen, Germany\\
}
\date{Accepted XXX. Received YYY; in original form ZZZ}
\pubyear{\the\year{}}

\begin{document}

	\label{firstpage}
	\pagerange{\pageref{firstpage}--\pageref{lastpage}}
	\maketitle

	\begin{abstract}
		Strong gravitational lensing can be used to find otherwise invisible dark matter subhaloes. In such an analysis, the lens galaxy mass model is a significant source of systematic uncertainty. In this paper we analyse the effect of angular complexity in the lens model. We use multipole perturbations which introduce low-order deviations from pure ellipticity in the isodensity contours, keeping the radial density profile fixed. We find that, in HST-like data, multipole perturbations consistent with those seen in galaxy isophotes are very effective at causing false positive substructure detections. We show that the effectiveness of this degeneracy depends on the deviation from a pure ellipse and the lensing configuration. We find that, when multipoles of one per cent are allowed in the lens model, the area in the observation where a subhalo could be detected drops by a factor of three. Sensitivity away from the lensed images is mostly lost. However, the mass limit of detectable objects on or close to the lensed images does not change. We do not expect the addition of multipole perturbations to lens models to have a significant effect on the ability of strong lensing to constrain the underlying dark matter model. However, given the high rate of false positive detections, angular complexity beyond the elliptical power-law should be included for such studies to be reliable. We discuss implications for previous detections and future work.

	\end{abstract}


	
	\section{Introduction}
\label{sec:introduction}
Dark matter is assumed to dominate the matter density of the Universe, yet, its exact nature remains unknown \citep{Planck2020}. The most popular theory, cold dark matter (CDM) proposes the existence of an as-yet undetected massive particle. In this paradigm, dark matter structures, called haloes, build up hierarchically in a self-similar way \citep{Springel2008,Wang2020}. The largest haloes host galaxies, but also an abundance of smaller subhaloes which may not hold any baryonic component. Warm dark matter (WDM) models have a smaller particle mass, leading to a suppression in the formation of subhaloes below a certain mass scale, called the half-mode mass $\Mhm$, inversely proportional to the particle mass \citep{Lovell2014}. Other dark matter models also make testable predictions for dark matter features on a sub-galactic scale, e.g., self-interacting dark matter \citep[SIDM,][]{Vogelsberger2012}, and fuzzy dark matter \citep[FDM,][]{May2023}. Finding such substructures, or not, in the Universe and characterising their properties therefore constrains the real dark matter model.

In pursuing that goal, strong gravitational lensing has emerged as one of the most promising tools \citep[][and references therein]{Vegetti2023}. Strong lensing observations are sensitive to the distribution of mass in the lensing galaxy's environment, and along the line of sight between source and observer. It can therefore be used to infer the presence, or absence, of dark matter haloes, regardless of their baryonic content. A number of methods, which we summarise here, exploit this fact to constrain dark matter models.

In gravitational imaging the observation is first fit with a model for the main lensing galaxy's mass distribution \citep{Vegetti2009}. Pixelated corrections to the lensing potential are then found which improve the fit to the data, up to some regularisation conditions. The corrected density map is then checked for overdensities consistent with low-mass haloes, which can be fit parametrically. Two low-mass haloes have been detected in this way \citep{Vegetti2010,Vegetti2012}. Other authors have adopted a method that uses only parametric fitting to find low-mass haloes, finding results consistent with gravitational imaging in the same data \citep{Sengul2022,Nightingale2022,Ballard2023}. Both gravitational imaging and parametric methods rely on lensed images of extended sources, typically in Einstein ring-like configurations. Other authors have instead used lensed quasars which appear as point sources. Anomalies in the flux ratios of these images can be used to place constraints on the bulk properties of dark matter, although the properties of individual objects are not constrained \citep[e.g.][]{Gilman2020,Hsueh2020}.

In this paper we examine one significant source of systematic error common to all above methods, the possible angular complexity in the main lens galaxy mass model. We show that small changes in the angular structure of the lens galaxy can easily be mistaken for substructures when an insufficiently complex model is used. When a more complicated model is used, the area of the observation in which a substructure can be detected drastically changes, although the smallest detectable mass of substructure does not. We argue therefore that angular complexity is required in strong lensing models for dark matter constraints to be reliable. Indeed, this statement likely applies to many other methods that rely on accurate strong lens modelling.

The most commonly used model for strong lensing analyses of any kind, not just that relating to substructure detection, is the elliptical power-law, or some form of it \citep{Tessore2015,ORiordan2020b}. In this model the projected total mass density $\kappa$ decreases with radius $\theta$ with some slope $\gamma$, i.e., $\kappa\propto\theta^{-\gamma}$. The slope is occasionally fixed to $\gamma=2$, in which case it is called a singular isothermal ellipsoid (SIE). Projected isodensity contours are similar ellipses with axis ratio $q$ and the major axis points in the direction $\phi_\mathrm{L}$. There are then two angular degrees of freedom and one radial. Other models add radial complexity by allowing for a change in slope \citep[broken power-law,][]{ORiordan2021} or decomposing the total mass profile into separately modelled baryonic and dark matter components where the lens galaxy light informs the fit to the baryonic part \citep{Nightingale2019}.

If extra angular degrees of freedom are included, this is typically by the addition of multipole perturbations. These are formulated as Fourier modes, where higher order perturbations describe smaller scale deviations from perfect ellipticity in the galaxy. Multipole perturbations or similar are often used in modelling \citep[e.g.][]{Evans2003,Gomer2018,Gomer2021,Powell2022,Nightingale2022,Gilman2023,Etherington2023} but their effect on substructure detection has not been systematically studied. \citet{VandeVyvere2022a} has studied their effect on time-delay cosmography. In all cases, multipole amplitudes of order 1 per cent are typically found when fitting to strong lensing data.

The formulation and use of multipoles is well motivated by observations. The isophotes of elliptical galaxies are consistently found to deviate from ellipses in a way that is well fit by Fourier modes, especially the $m=4$ mode which can describe boxy/discy features \citep{Bender1989}. \citet{Hao2006} find deviations are strongest in the $m=4$ mode, with amplitudes of $\sim1$ per cent while perturbations of order three are typically smaller (see their fig. 4). At galactic radii larger than $~1.5$ times the half-light radius, these perturbations can be much stronger \citep{Chaware2014}. In the redshift range typically occupied by strong lens galaxies, there is no systematic evolution in the multipole parameters \citep{Mitsuda2017,Jiang2018}. In simulations of elliptical galaxies, the pre-merger mass-ratio is found to determine the isophote shape in the remnant \citep{Naab2003}. Isophotes are known to exhibit other non-elliptical features not described by multipoles such as twists and ellipticity gradients \citep{VandeVyvere2022b}, but those features are not studied here. Informed by these results from observations and simulations, we consider in this work multipole perturbations up to order four and up to strengths of one and three per cent.

The paper is organised as follows. In \cref{sec:method} we describe the mock observations we use as well as the method for detecting subhaloes and producing sensitivity maps. In \cref{sec:results} we present our results. In \cref{sec:discussion} we discuss the implications of the results. Finally, in \cref{sec:conclusions} we summarise our conclusions. Throughout this paper we assume a Planck 2015 cosmology with $H_0=67.7\,\mathrm{kms}^{-1}\mathrm{Mpc}^{-1}$ and $\Omega_\mathrm{m,0}=0.302$ \citep{Planck2015}.

	\section{Method}
\label{sec:method}

In this section and later we frequently refer to a previous paper defining the underlying method for substructure detection and sensitivity mapping, \citet{ORiordan2023}, hereafter referred to by \citetalias{ORiordan2023}.

We refer throughout the paper to data in different ways, depending on its use. First, training data are those used only to train our machine learning substructure detectors. These data are realistically simulated as described below, but certain parameters are resampled from uniform distributions when the images are realised. This improves the generality of the detector model. These training data can include angular complexity, depending on the model being trained. Second, a smaller set of testing data is produced in the same manner as training data. This is used to validate model performance during training. Third, there is scientific data which is used after training to create all of the scientific results in \cref{sec:results}. In these data, the realistically simulated properties of the observation are preserved, i.e., they are realised exactly as simulated. These data do not contain any multipole perturbations so as to avoid complicated interactions between substructure and multipoles in the data, rather than in the model. The differences are summarised in \cref{table:data}.

\begin{figure*}
	\includegraphics[width=1.0\textwidth]{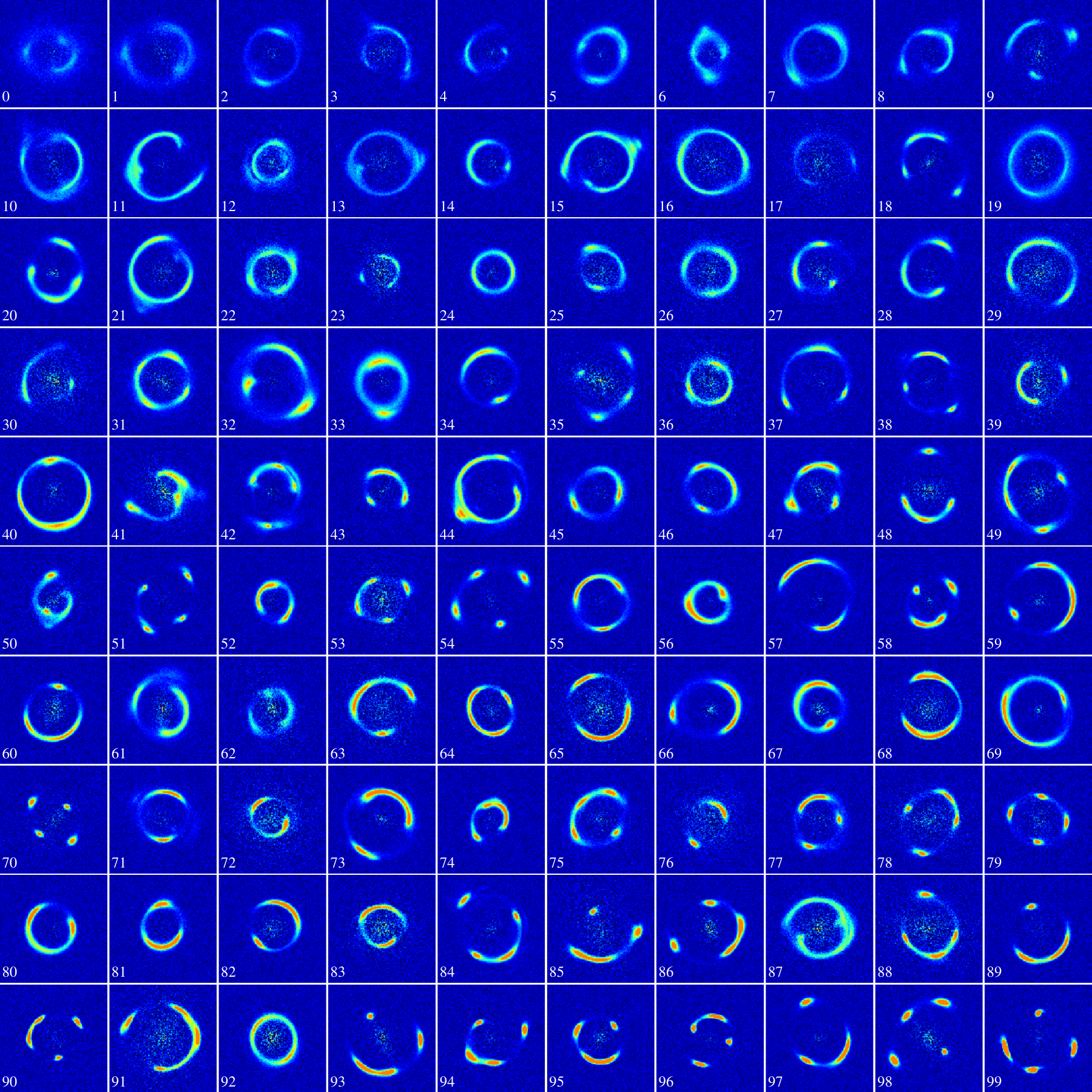}
	\caption{\label{fig:lens-gallery}HST mock images used in this work. Images are ordered by peak brightness and all images share the same scale. The S\'ersic profile for the lens galaxy light has been subtracted but Poisson noise from this emission remains in the centre. These data are not representative of the training data for our models, which use a much larger space of parameters and S/N, see \cref{sec:training}.}
\end{figure*}

\begin{figure*}
	\includegraphics[width=1.0\textwidth]{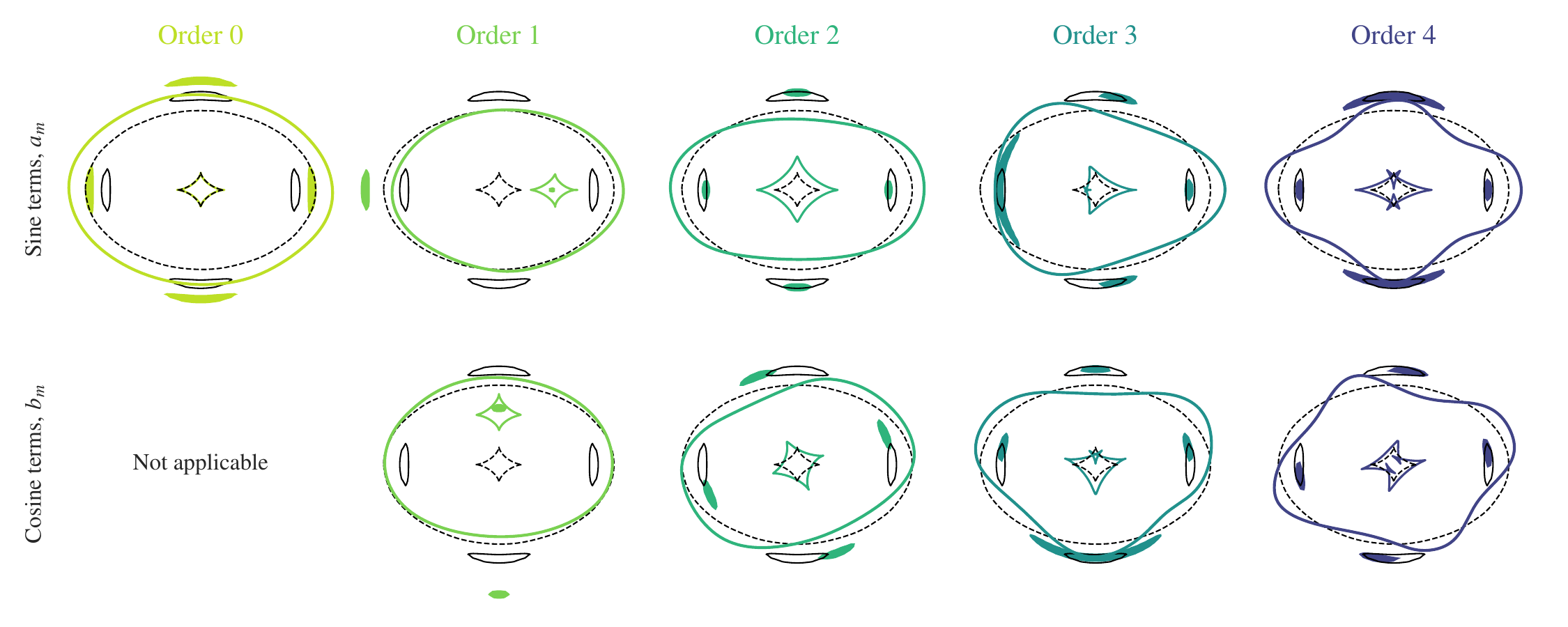}
	\caption{\label{fig:multipole-critical-curves} Changes to the critical curves, lensed images, and caustics due to multipole perturbations in an isothermal lens with axis ratio $q=0.7$ and a source at the origin. The unperturbed critical curves and caustics are plotted as dashed black lines. The perturbed versions are plotted as solid coloured lines. In each case the perturbation is $0.1$ in the respective coefficient. This is an extremely large amplitude and is only used for illustration purposes. The monopole ($m=0$) perturbation is isotropic and only has one coefficient. It is identical to a change in the lens Einstein radius and so is not used in this work. Similarly, the quadrupole ($m=2$) perturbations replicate a change in the lens axis ratio or position angle, and so are also not included.}
\end{figure*}

\subsection{Mock data}

We generate training and testing data in the same step. Our scientific data are drawn from the same simulated catalogue of lens and source parameters used for the testing data, but the multipole perturbations are excluded. For this scientific data, we use a simulated set of $100$ Hubble Space Telescope (HST) optical images. All results in \cref{sec:results} use these mock observations. They are intended to be representative of data previously used for gravitational imaging.

We draw the mock observation parameters from a catalogue of simulated strong lens systems created in \citetalias{ORiordan2023}, which should be consulted for details. To summarise the simulation procedure in that paper: lens galaxies are assumed to be described by elliptical power-law density distributions, with a mass, axis ratio, and redshift drawn from appropriate realistic distributions for elliptical galaxies in the Local Universe. Their mass distributions do not contain dark matter substructure, or angular features beyond the ellipticity of the power-law, and an external shear. They are placed in the foreground of Hubble Deep Field (HDF) galaxies which are to be used as lensed sources. The light cone behind the lensing galaxy is checked for sources which will form multiple images in the lens plane. So as not to imprint the real redshift distribution of the sources on the training data, the original source redshift is perturbed by resampling a redshift close to the original. The lens galaxy apparent magnitude, size, and S\'ersic index are drawn from appropriate scaling relations. The combination of lens and source properties are then catalogued in batches of $10^5$ lens-source systems each. $40$ batches are produced for training, and $10$ batches for testing data.

\begin{table}
	\begin{tabular}{ l l l }
		Data type & Number of images & Multipole perturbations included\\
		\hline
		Training data & $3\times4\times 10^6$ & In $\modeltwo$ and $\modelthree$, not in $\modelone$\\
		Testing data & $3\times 10^6$ & In $\modeltwo$ and $\modelthree$, not in $\modelone$\\
		Scientific data & $10^2$ & None\\
		\hline
	\end{tabular}
	\caption{\label{table:data} The three different types of data used in this paper, the number of images in each, and whether multipole perturbations are present in the lens model. For training and testing data, there are a further three subsets, explained in \cref{table:models}. The scientific data do not contain multipole perturbations when created, but they are effectively modelled with multipoles whenever $\modeltwo$ or $\modelthree$ are used.}
\end{table}

To select our scientific data, we draw all systems from the first testing catalogue which satisfy the following conditions,
\begin{equation}
	\label{eq:conditions}
	\begin{split}
		&0.8\leq\erad/\mathrm{arcsec}\leq1.6,\\
		&\beta/\erad < 0.2,\\
		&q>0.5,\\
		&50<\mathrm{S/N}<300,
	\end{split}
\end{equation}
where $\erad$ is the Einstein radius, $\beta$ is the radial source position, $q$ is the axis ratio of the lens projected mass distribution, and $\mathrm{S/N}$ is the total signal to noise ratio in the lensed images. The conditions are intended to produce systems with large and full Einstein rings. This type of system is most favourable for gravitational imaging due to the large area on the sky where a low-mass perturber could sufficiently disrupt the lensed emission to be detected. The conditions also produce systems that mimic those in the SLACS and BELLS GALLERY strong lens surveys \citep{Bolton2008,Brownstein2012}. These same lenses have been the subject of many strong lensing substructure studies performed so far \citep[e.g.][]{Vegetti2012,Ritondale2019,Nightingale2022}.

From the original catalogue of $10^5$ lens-source pairs, $1798$ satisfy the conditions in \cref{eq:conditions}. From these we randomly choose $100$ systems which form the dataset used in this paper. Although the catalogue was originally created to simulate systems for Euclid, here we produce images with resolution, PSF, and noise for the HST Wide Field Camera F606 filter, i.e., with a pixel size of $0.04$ arcsec and a Gaussian PSF with FWHM of $0.07$ arcsec. The mock observations are plotted in \cref{fig:lens-gallery}. Note that the source surface brightness distributions  obtained in \citetalias{ORiordan2023} and used here are a denoised combination of images from three WFC bands: F606, F775, and F814, whereas SLACS and BELLS images are in F814 only. There may be therefore slight differences in source structure although visual inspection of the figure shows this to be minimal.

\subsection{Model training}
\label{sec:training}
Our method for substructure detection relies on the machine learning apparatus developed in \citetalias{ORiordan2023}. We restate some of the main features of that work here. We use a deep convolutional neural network which predicts the presence of substructure in strong lens observations. The architecture is a 50 layer residual neural network \citep[ResNet]{He2015}. Each training example is a simulated strong lens observation which contains either zero or between one and four substructures. The lens and source parameters from the simulated catalogues in the previous section are combined with substructure properties explained in this section to realise an image. For a given image, the neural network returns the probability that a non-zero amount of substructure is in the image.

We produce $4\times10^6$ training images, with a tenth that number used for testing. When producing the training data, the macro properties of each lens-source system, simulated according to the previous section, are replaced with properties drawn from uniform distributions, if a given parameter allows. For example, rather than using the realistic distribution of Einstein radii in our simulations, which has a heavy tail with a peak around $0.7$ arcsec, we draw uniformly from $0.5\leq \erad/\mathrm{arcsec}\leq 2.0$. This improves the generality of the neural networks as specific configurations and data conditions which are common in nature are not over-represented in the training data. The training data can therefore be imagined as the space of all possible lenses, sources and S/N, while the data used in the results are a subset which is realistic.

In the training data, the substructures have maximum velocities $\vmax$ drawn log-uniformly from the range $10\,\mathrm{kms}^{-1} \leq \vmax \leq 158\,\mathrm{kms}^{-1}$. This is roughly equivalent to the mass range $10^7\leq\Mmax/\Msun\leq10^{11}$.\footnote{This mass range is similar to that of the Milky Way satellite galaxies, so objects of this size may well have a luminous component. However, at the distance of a typical strong lens, they are only detectable gravitationally.} The subhaloes use Navarro-Frenk-White (NFW) profiles with $\rmax$ drawn log-uniformly from the range $1\leq \rmax/\mathrm{kpc}\leq28$. In our sensitivity maps and our results the subhalo $\rmax$ is instead drawn from the redshift dependent $\vmax$-$\rmax$ relation of \citet{Moline2022}. It should be noted that this relation is more accurate for subhaloes, and produces objects more concentrated than the traditionally used relation of \citet[][see also Table 1 of \citetalias{ORiordan2023}]{Duffy2008}. In this work we only consider subhaloes, i.e., perturbers in the plane of the main lens. It is not straightforward to apply the models described to field haloes, i.e., those along the line of sight between observer and source, due to complicated effects introduced by multiplane lensing. This will be examined in a dedicated future paper. The lensed source surface brightness distributions are taken from images of HDF galaxies, with separate sets used for training and testing data. The testing set of source images are also those used in our results.

We train the model in stages where the detection problem becomes more difficult in each stage, by e.g. broadening the S/N range or adding more complex model features like external shear. This allows us to monitor the response of the model to specific additional complexities, and helps speed up convergence. We minimise the cross-entropy loss of the model using the Adam optimizer. After a parameter sweep to find the best learning rate, training continues with the learning rate decayed by a factor $10^{-0.25}$ every time the test loss does not improve for $10$ epochs. After three consecutive learning rate decays with no test loss improvement the model is assumed to have converged.

\begin{table}
	\begin{tabular}{p{0.5cm} p{0.89cm} p{0.89cm} p{0.89cm} p{0.66cm} p{0.62cm} p{0.89cm} }
		Model & $\erad$ & $q$ & $\gamma$ & $\abs{\gamma_\mathrm{ext}}$ & $m$ & $\mstrength$\\
		\hline
		$\modelone$ & 0.5 - 3.0 & 0.4 - 1.0 & 1.8 - 2.2 & $<0.1$ & None & 0.0\\
		$\modeltwo$ & 0.5 - 3.0 & 0.4 - 1.0 & 1.8 - 2.2 & $<0.1$ & 1,3,4 & $<0.01$\\
		$\modelthree$ & 0.5 - 3.0 & 0.4 - 1.0 & 1.8 - 2.2 & $<0.1$ & 1,3,4 & $<0.03$\\
		\hline
	\end{tabular}
	\caption{\label{table:models}The range of parameters in the training data for each of the three models. In effect, these are the prior probability distributions for each parameter. The columns are; the Einstein radius $\erad$ in arcsec, the axis ratio of the elliptical lens mass profile $q$, the slope of the lens power-law mass profile $\gamma$, the strength of the external shear component $\gamma_\mathrm{ext}$, the order of allowed multipoles $m$, and the strength of the allowed multipole perturbations $\mstrength$. For $\modeltwo$ and $\modelthree$ the multipole angle $\mangle$ is drawn uniformly between $0$ and $2\pi$.}
\end{table}

\subsection{Internal angular structure}
\label{sec:method-multipoles}
We use three distinct models for substructure detection which vary in the allowed angular complexity for the macro model. This is achieved by training the three models on similar but separate datasets. Each model then effectively learns a different prior for the macro model. In the first model, $\modelone$, the lens galaxy mass profile is an elliptical power-law plus external shear. The lens Einstein radius, axis ratio, mass profile slope, position angle, and external shear strength and angle vary across a large uniform range in the training data. The second and third models, $\modeltwo$ and $\modelthree$, use the same ingredients as $\modelone$ but add multipole terms to the convergence of the lensing galaxy. We use the definition given by \citet{Powell2022} for the convergence $\kappa_m$ due to a multipole of order $m$,
\begin{equation}
	\kappa_m(\theta,\phi)=\theta^{1-\gamma}\left[a_m\sin(m\phi) + b_m\cos(m\phi)\right],
\end{equation}
where $\theta$ is the angular radial distance from the centre of the lens, $\phi$ is the polar angle, $\gamma$ is the slope of the lensing galaxy power-law mass profile, and $a_m$ and $b_m$ are the sine and cosine amplitudes. The slope of the multipole components is the same as the power-law for the sake of simplicity, and so as not to introduce extra radial complexity which might have an effect beyond the angular complexity we focus on here. In general, there may be scenarios which require a separate slope for the multipole component. We also define the multipole strength, $\mstrength$, and multipole angle, $\mangle$, as
\begin{align}
	\mstrength^2&=a_m^2+b_m^2,\\
	\mangle&=\arctan(b_m/a_m),
\end{align}
respectively. \Cref{fig:multipole-critical-curves} shows the effect of these multipole perturbations on the lens.

Models $\modeltwo$ and $\modelthree$ use multipoles of order 1, 3, and 4. The quadrupole ($m=2$) perturbation is not used in any model because the elliptical lens mass model and the external shear already have this complexity. In the training data for $\modeltwo$, $\eta_m$ is drawn uniformly in the range $\eta_m<0.01$. In $\modelthree$ this is extended to $\eta_m<0.03$. The multipole angle $\phi_m$ is drawn from $0\leq\phi_m<2\pi$. The coefficients for each order are drawn independently, so that training examples can include any combination of the three orders. The models are summarised in \cref{table:models}.

\subsection{Sensitivity maps}
\label{sec:sensitivity-maps}
A sensitivity map gives the lowest detectable mass of substructure at every position on the observation, given some detection threshold. The sensitivity depends also on the choice of profile for the substructure, and its concentration. Both are fixed in this case.
The trained neural network can be thought of as a black box substructure detector $S(d)$ that returns the posterior probability $\prob{\mathrm{sub}}{d}$ for a given observation $d$. The prior implicit in this posterior is the prior in the training data itself. As such, any degeneracy or complexity in the training data is accounted for by this probability. For example, when external shear is added to the training data, which is in effect an $m=2$ multipole perturbation and degenerate with substructure, the detection significances reported by the model drop for all substructure masses \citepalias[see][fig. 3]{ORiordan2023}.

The detector can be used to construct a sensitivity map as follows. For a given lens-source system, introduce a test subhalo in a position $(\theta_x,\theta_y)$ with mass and concentration parametrised by $\Mmax$ and $\rmax$. Ray-trace the system to produce an image $d$. Keeping the properties of the lens, source, and noise realisation fixed, the image is just a function of the subhalo properties. We move the test mass across the image pixel grid and iterate over a range of interesting masses, in this case $10^{7.6}<\Mmax/\Msun\leq10^{11}$ with $35$ steps in mass spread log-uniformly and $\rmax$ is set to the mean of the relation at the redshift of the lens. The detector then returns the probability that the substructure is in the image, as a function of its properties, i.e., $S(\theta_x,\theta_y,\Mmax,\rmax)$. In practice we convert this to a more conventional detection significance.

\subsection{Expected number of detections}
\label{sec:method-expected-number}
The number of subhaloes $\dd{n}$ of mass $m$ in a mass interval $\dd{m}$ per projected area on the sky is proportional to the subhalo mass function
\begin{equation}
	\label{eq:mass-function}
	\derivative{n}{m} \propto m^{\alpha_1}\left[1+\left(\alpha_2\frac{\Mhm}{m}\right)^\beta\right]^{\gamma},
\end{equation}
where the constants have the values; $\alpha_1=-1.9$, $\alpha_2=1.1$, $\beta=1.0$, and $\gamma=-0.5$. These are obtained from the simulations in \citet{Lovell2020b}, re-fit to be in terms of $\Mmax$ \citepalias{ORiordan2023}. The half-mode mass $\Mhm$ parametrises the suppression of low-mass structures in WDM, and is the mass scale where the WDM power-specturm is half of that in CDM. For a lens galaxy with a mass enclosed by twice the Einstein radius, $\Menc$, the expected number of detectable subhaloes in the $i$th pixel is
\begin{equation}
	\label{eq:mu-integral}
	\mu_{\mathrm{det}_i}=A_i\fsub^\mathrm{CDM}\frac{\Menc}{\Mtot} \displaystyle\int_{\mmap}^{\mmax}\dv{n}{m}\dd{m},
\end{equation}
where $A_i$ is the ratio of the pixel area, $\omega_i$ to the area that defines $\fsub$, in this case,
\begin{equation}
	A_i=\frac{\omega_i}{4\pi\erad^2}.
\end{equation}
The total mass in substructure in CDM, $\Mtot$, is given by
\begin{equation}
	\label{eq:norm-integral}
	\Mtot=\int_{\mmin}^{\mmax}m^{\alpha_1+1}\dd{m}.
\end{equation}
Using a CDM mass function to normalise a general mass function in this way ensures that warmer models do not gain larger numbers of substructures for the same normalisation. The overall number is always suppressed in warmer models, as well as at lower masses. Note that \cref{eq:mu-integral} and \cref{eq:norm-integral} have different integration limits. In 
\cref{eq:norm-integral}, $\mmin$ and $\mmax$ are the lower and upper limits of all the substructure mass that we consider, regardless of whether they are detectable. In \cref{eq:mu-integral}, $\mmap$ is the lowest detectable mass returned by the sensitivity map in that pixel. Finally, the expected number of detectable subhaloes in an observation, $\mu_\mathrm{det}$ is just the sum of $\mu_{\mathrm{det}_i}$ over all pixels with $\theta<2\erad$.

	\section{Results}
\label{sec:results}

\begin{figure*}
	\includegraphics[width=1.0\textwidth]{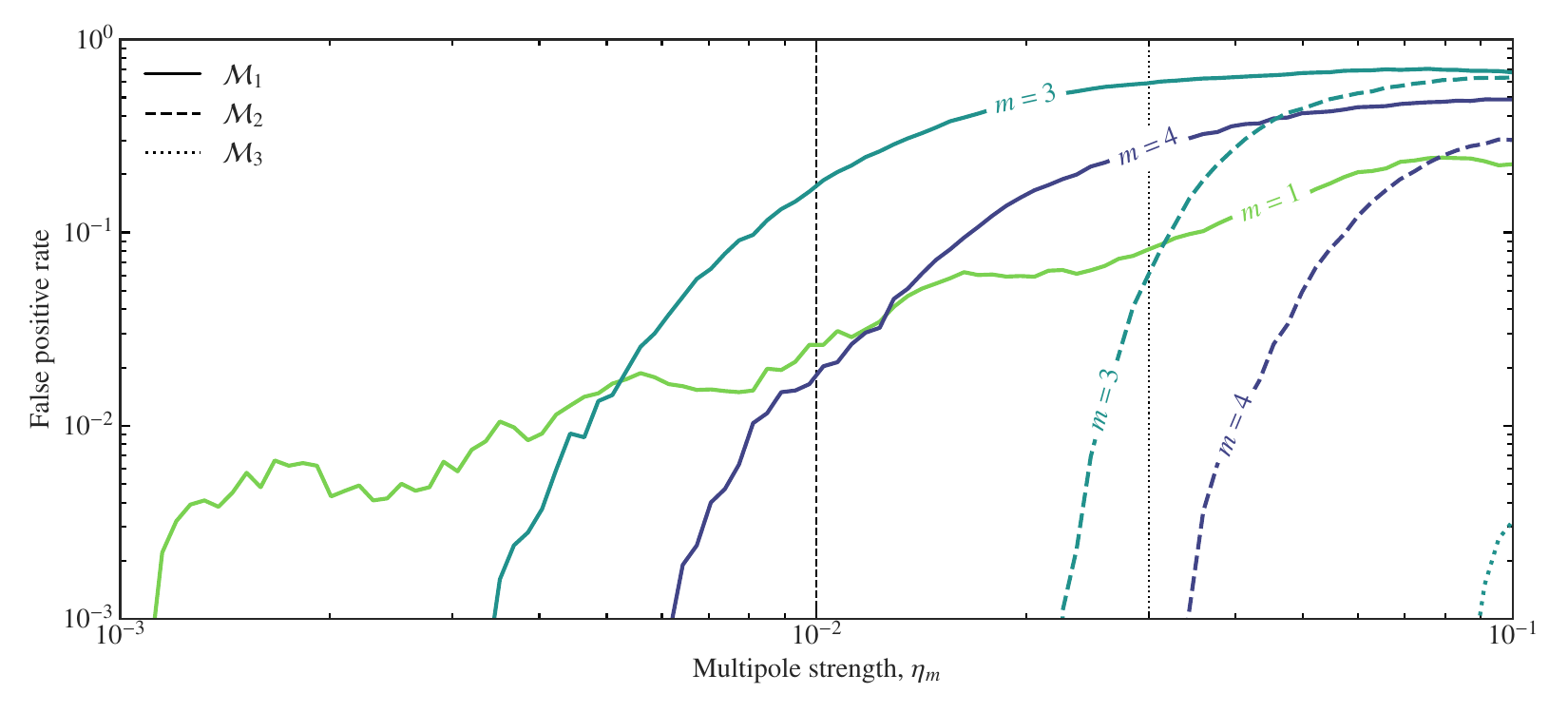}
	\caption{\label{fig:false-positives}False positive rate (FPR) of substructure detection in images containing multipole perturbations and no substructures, as a function of multipole strength $\mstrength$ and order $m$. Multipole order is labelled on each curve. Solid curves show the FPR for the model trained with no multipole perturbations. The dashed and dotted curves show the FPR for the models with prior ranges $\eta_m<0.01$ and $\eta<0.03$ respectively, also indicated by vertical lines.}
\end{figure*}

\begin{figure*}
	\includegraphics[width=1.0\textwidth]{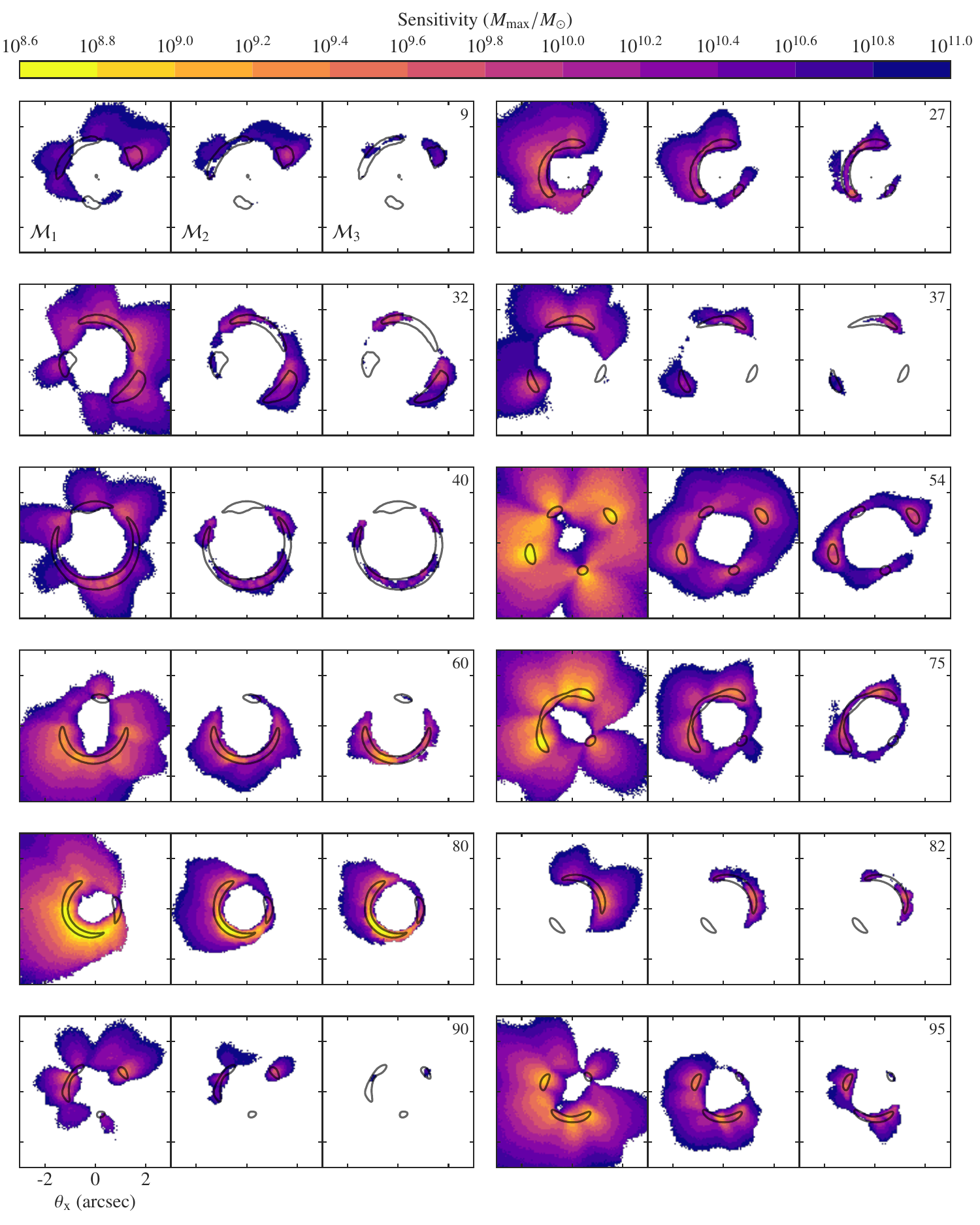}
	\caption{\label{fig:sensitivity-maps} A sample of 12 sensitivity maps for our mock HST images. The maps show the lowest mass subhalo detectable at $5\sigma$ significance for the same observation using models $\modelone$, $\modeltwo$, $\modelthree$. The left frame for each lens is then the sensitivity if multipoles are not included in the lens model. The centre and right frames show the sensitivity when multipoles of strengths $\mstrength<0.01$ and $\mstrength<0.03$ are included in the model respectively. Black contours overlay the extent of the lensed images. The systems are numbered in the same way as in \cref{fig:lens-gallery}, so that figure can be checked for the corresponding observations.}
\end{figure*}

\begin{figure*}
	\includegraphics[width=1.0\textwidth]{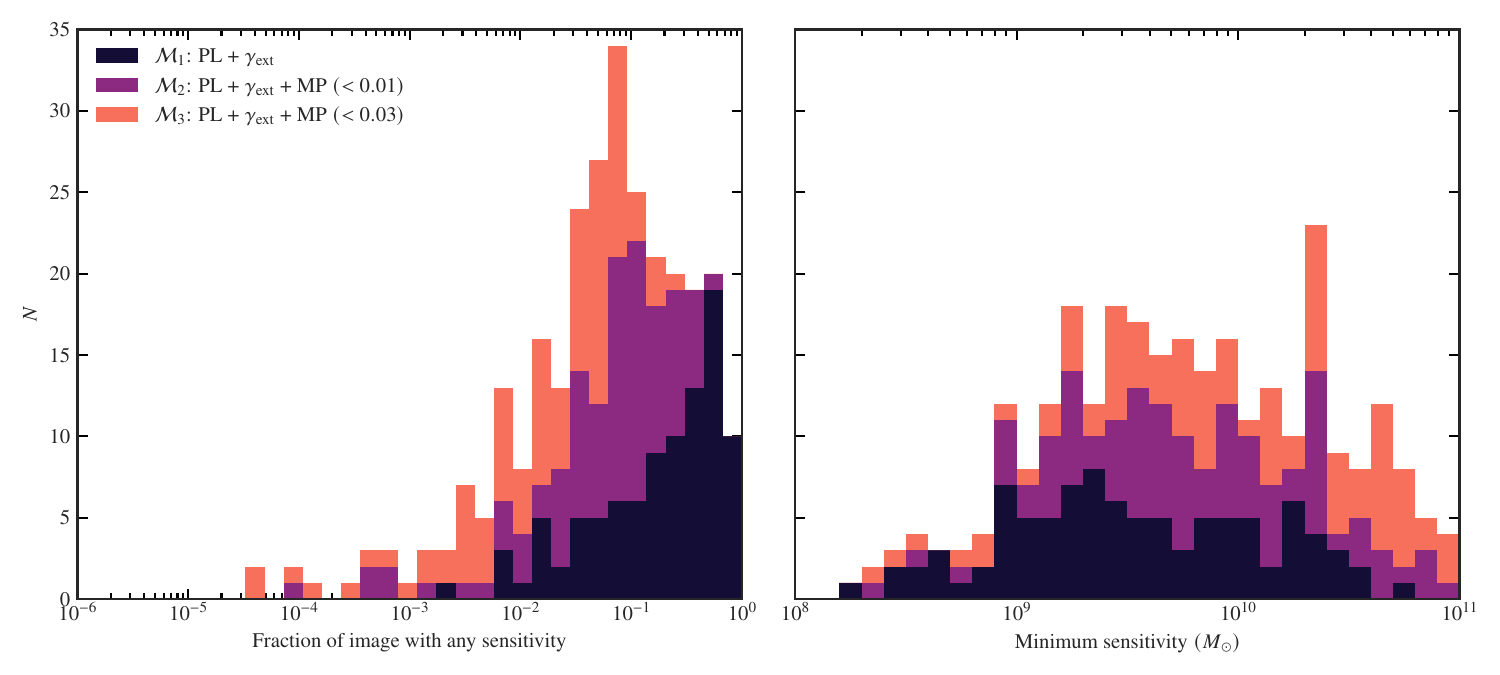}
	\caption{\label{fig:sensitivity-statistics} Distributions of two sensitivity statistics for all $100$ mock HST observations. In both frames the differently coloured samples are stacked on top of each other. Left: the fraction of the observation which is sensitive to any mass of substructure $\Mmax<10^{11}\Msun$ at $5\sigma$. Right: the detectable mass in the most sensitive pixel in each observation. Different colours indicate the different models. The sum over all $N$ is not necessarily the total number of observations ($100$) because some systems lack any sensitivity at all in the tested range.}
\end{figure*}

\begin{figure}
	\includegraphics[width=1.0\columnwidth]{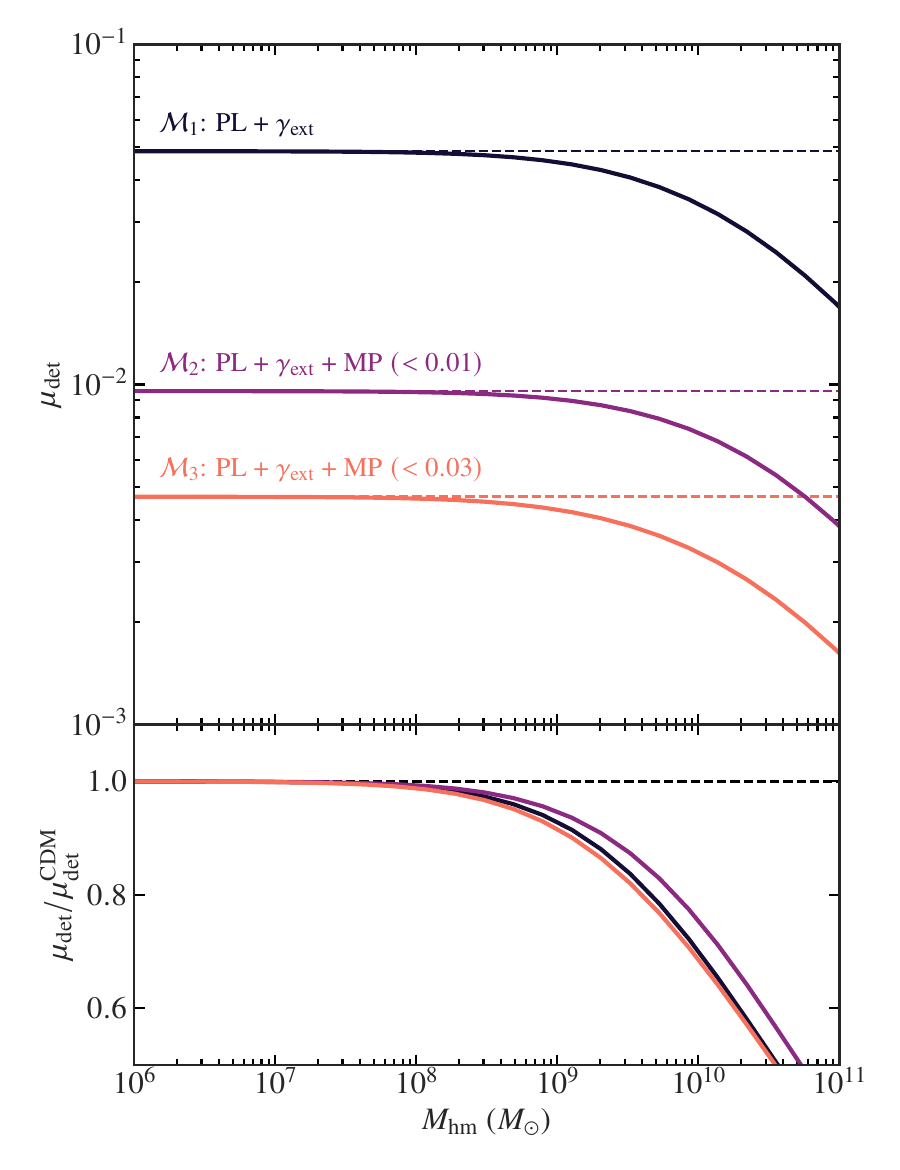}
	\caption{\label{fig:mass-function-statistics} Upper frame: the expected number of detections per lens, $\mu_\mathrm{det}$, in $100$ HST mock observations, using sensitivity maps from three different models. Lower frame: the expected number of detections per lens relative to that in CDM, $\mu_\mathrm{det}^\mathrm{CDM}$. The detection rate in both frames is plotted as a function of the half-mode mass $\Mhm$. Horizontal dashed lines give the CDM value.}
\end{figure}

\subsection{False-positive substructure detections}
\label{sec:false-positives}
We first demonstrate that multipole perturbations are easily mistaken for substructures by $\modelone$, the model restricted to using an elliptical power-law plus external shear for the main lens. To do this, we create new realisations of our scientific data, the 100 mock HST lenses, and now include multipole perturbations in the lens model. For a single multipole order $m=\{1,3,4\}$, we sample $100$ multipole strengths log-uniformly in the range $10^{-3}\leq\mstrength\leq10^{-1}$. At each value we create $100$ realisations of each of the $100$ mock systems in \cref{fig:lens-gallery}, with randomly sampled multipole angles $\mangle$ in the range $0\leq\mangle<2\pi$. We pass each realisation through $\modelone$ and record the rate of $5\sigma$ detections. For completeness we also pass these realisations through $\modeltwo$ and $\modelthree$, although we do not expect these models to produce false positive detections within their prior range of $\mstrength$.

\Cref{fig:false-positives} shows the false positive rate (FPR) of substructure detection as a function of multipole strength $\mstrength$ and order $m$ for the three models. Consider first the model with no multipole perturbations in the training data, $\modelone$. The figure shows that modest perturbations cause false positive substructure detections at a significantly high rate. A perturbation of $\mstrength=0.01$ is detected as a substructure 2.6 per cent, 18 per cent, and 2.0 per cent of the time for orders one, three, and four respectively. At $\mstrength=0.03$ these same FPRs are 9.4 per cent, 61 per cent, and 30 per cent. At all multipole strengths, the order three perturbation is the most effective at causing false positive detections, with order four having a similar but weaker effect.

The figure also indicates the prior ranges for $\mstrength$ in the two models which include multipoles in the training data, $\modeltwo$ and $\modelthree$. In both cases, it is clear that multipole perturbations only cause false positive detections with strengths well outside of the prior range. A non-zero FPR is only recorded for $\modeltwo$ when $\mstrength>0.023$, and for $\modelthree$ when $\mstrength>0.079$, more than twice the size of the prior in both cases. Both of these models have therefore successfully learned the degeneracy between angular structure in the lens, and dark matter substructures, within their respective prior ranges.

\subsection{Sensitivity mapping with multipoles}

Using the procedure described in \cref{sec:sensitivity-maps}, we compute sensitivity maps using all three models for our $100$ mock lenses. These scientific data are now used in their originally simulated state, i.e., they do not contain multipole perturbations. \Cref{fig:sensitivity-maps} shows these maps for $12$ of the $100$ systems. The figure shows the lowest mass subhalo that each of the three models could detect in each pixel of the observations at our chosen detection significance of $5\sigma$.

First consider the $\modelone$ sensitivity maps alone. This gives us an indication of the sensitivity of HST observations in general, without considering the effect of multipole structure. Of the $100$ systems, $18$ have at least one pixel sensitive below $\Mmax=10^{9}\Msun$. The most sensitive pixel in the sample is on the main arc of the system labelled `80', where a subhalo with mass $\Mmax=10^{8.2}\Msun$ can be detected at $5\sigma$. Now consider the change in sensitivity as we add multipoles to the model, comparing $\modelone$ and $\modeltwo$. The ground truth mock observations do not contain multipole perturbations in the lens model, so any difference in the sensitivity maps between the models is due to the similar effect that both dark matter substructures and angular structures in the lens galaxy have on the lensed images.

The general effect of this degeneracy is to reduce the area in the observation where any substructure could be detected, removing the large sensitive areas surrounding the lensed images. However, the level of sensitivity on the lensed images is only slightly reduced. The reduction in area is clearest in systems which start with a very large sensitive area, e.g., 54, 75, and 95. Comparing $\modeltwo$ and $\modelthree$, the increase in allowed multipole strength does not further degrade the sensitivity in the lensed images at all, but the sensitive area continues to reduce.

\Cref{fig:sensitivity-statistics} plots the distribution of two sensitivity summary statistics for all $100$ lenses. In the left frame, we see the same trend identified in \cref{fig:sensitivity-maps}, that is, allowing for angular complexity in the lens mass model drastically reduces the area in the image where a substructure can be detected. In $\modelone$, with no multipoles in the model, 29 per cent of all the image area in the sample is sensitive, but this drops to 10 per cent in $\modeltwo$ and 3.6 per cent in $\modelthree$. By contrast, the minimum sensitivity, shown in the right panel of \cref{fig:sensitivity-statistics} changes by very little. The minimum sensitivity is defined here as the sensitivity of the most sensitive pixel in the image. The best pixel in the sample loses only 0.11 dex of sensitivity when going from $\modelone$ to $\modeltwo$, with no further change in $\modelthree$

\subsection{Mass function statistics}

Using the method in \cref{sec:method-expected-number}, we can calculate the expected number of detectable objects in an observation, given its sensitivity map and a dark matter model. The dark matter model is described by just two parameters: $\fsub$, the fraction of mass contained in substructure, which normalises the mass function, and; the half-mode mass $\Mhm$, the mass below which structure formation is suppressed. In this work we always use a fixed $\fsub^\mathrm{CDM}=10^{-2}$, although the expected number of detections scales straightforwardly as $\mu_\mathrm{det}\propto\fsub^\mathrm{CDM}$ (Eq.~\ref{eq:mu-integral}).

\Cref{fig:mass-function-statistics} plots the expected number of detections per lens over all 100 mock HST observations, for sensitivity maps created with all three models. As before it is useful to consider the $\modelone$ case first, as this indicates the number of detections we expect in observations which are already well studied. In CDM we expect $0.048$ detections per lens, or one detection in every $\sim20$ lenses studied. Although this is a small number regime, this frequency is consistent with the frequency of detections in real HST data \citep{Vegetti2014,Nightingale2022}. The number of expected detections only differs significantly from CDM for values of $\Mhm$ which have already been ruled out, i.e., $\Mhm\gtrsim10^8\Msun$ \citep{Enzi2021}. This is a function of the sensitivity limit of the data, found earlier at $\Mmax=10^{8.2}\Msun$, itself primarily a function of the angular resolution. We discuss this further in \cref{sec:classical-comparison}.

In $\modeltwo$ and $\modelthree$, the expected number of detections per lens drops to $0.0095$ and $0.0047$ respectively, equivalent to a detection once in every $\sim100$ and $\sim200$ lenses. This is not to say that these models predict less dark matter substructure. Rather, by allowing for angular complexity in the lensing galaxy mass profile, objects that were previously detectable can now be equally well accounted for by modest multipole perturbations. As such, detections occur at lower or no significance and are more rare above the set threshold.

In the lower frame of \cref{fig:mass-function-statistics} we plot the change in $\mu_\mathrm{det}$ relative to CDM, as a function of $\Mhm$ for all three models. This quantity gives a sense of the constraining power for $\Mhm$ that each of the models has, independent of the choice of $\fsub$. In all three models, $\musub$ drops to 90 per cent of the CDM value at $\Mhm=10^{9.2}\Msun$.

The behaviour in both frames of \cref{fig:mass-function-statistics} follows directly from the corresponding frames of \cref{fig:sensitivity-statistics}. The reduction in sensitive area reduces the overall number of possible detections, but, these detections are mostly in regions of poor sensitivity, and would be subhaloes of a larger mass which do not provide useful constraints on $\Mhm$. The sensitive area which remains is on the lensed arcs and images, where the sensitivity is typically the best and the lowest possible masses can be detected.

The effect then, of adding multipole perturbations to the lens model is to make large mass subhaloes, far from the lensed images, more difficult to detect. Detections of lower-mass objects close to or on the lensed images remain just as feasible. Seeing as this second class of objects is the most informative for $\Mhm$, we do not expect the inclusion of lens angular complexity to significantly affect the ability of gravitational imaging studies to constrain dark matter models.
	\section{Discussion}
\label{sec:discussion}

Our results have implications for all efforts to constrain dark matter models with strong lensing data. We have shown that, at the least, multipole perturbations must be allowed in the lens for these efforts to be reliable. In this section, we discuss the way in which this might be done, as well as other questions arising from our results.

We do not discuss the limitations of the method which are specific to machine learning, e.g., reliance on the training data used, model architecture etc. These points are discussed in \citetalias{ORiordan2023} and the major components of the method remain unchanged. In fact, the greater diversity of training data in this work has likely gone some way to mitigating such issues. We also omit two major points of discussion from this work which are important considerations in low-mass halo detection, namely, the concentration of the objects and whether they exist as subhaloes or field haloes along the line of sight. A follow-up paper will describe precisely the relationship between sensitivity, concentration, and the redshift of the halo relative to the lens.

\begin{figure}
	\includegraphics[width=1.0\columnwidth]{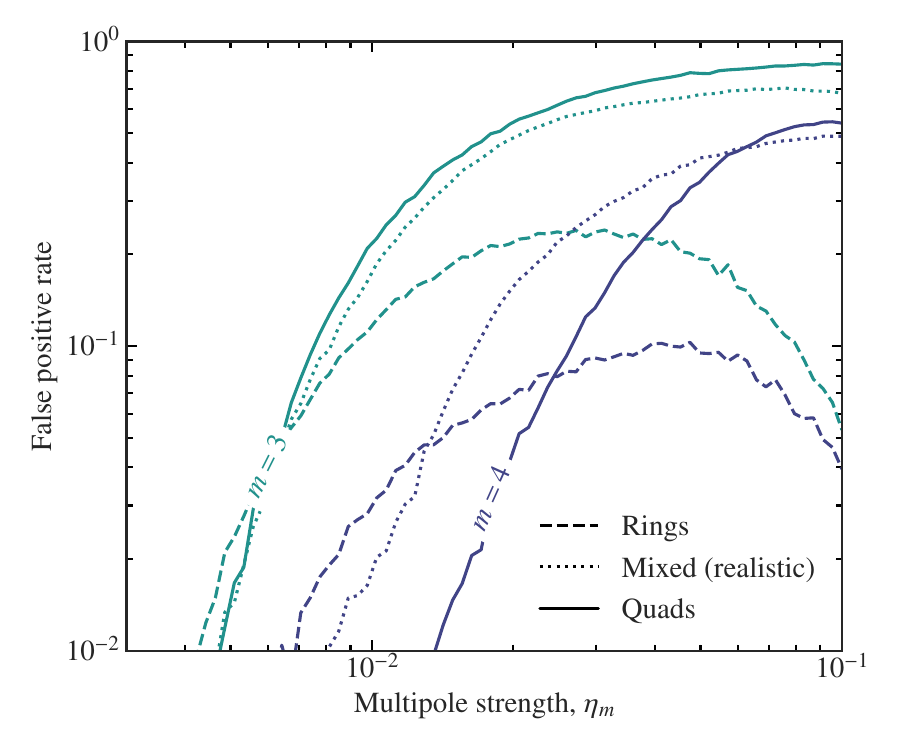}
	\caption{\label{fig:multipole-orders} False positive rate of substructure detection in $\modelone$ as a function of multipole strength and order for images without substructures. Three samples of images are shown: rings (dashed), quads (solid), and `mixed' (dotted). The third sample is that taken from realistic simulations and used previously, also shown in \cref{fig:lens-gallery}. The multipole orders are similarly coloured between the samples.}
\end{figure}

\subsection{Dependence on lensing configuration}

In \cref{sec:false-positives}, we showed that multipole perturbations in general can cause a significant rate of false positive substructure detections. \Cref{fig:false-positives} shows that this depends exactly on the order and strength of the perturbation. Somewhat counter-intuitively, order $3$ perturbations are more effective at causing false positive detections than order $4$. This is due to an interaction between the configuration of the lensed images, the multipole perturbations, and the substructure detector.

Our substructure detector models expect to see localised changes to the lensed images which are not consistent with changes to the macro model, e.g., there should only be a change in one image, or in a small part of the arc. For a multipole perturbation to cause a false positive detection, it must cause a similarly local change. This is not always straightforward however, as multipoles are global perturbations. Only when they align with the image configuration so that only one part of the arc, or one image, is perturbed, a false detection is possible. Here we show that such a scenario is much easier to achieve with order three perturbations than order four.

If the above is true, then the best case lensing configuration for mitigating the multipole degeneracy is one with a very high degree of symmetry, i.e., a full Einstein ring. At the other extreme, the worst case scenario is one with a low degree of symmetry, i.e., quads or doubles. To test this, we repeat the exercise in \cref{sec:false-positives} with two lensing samples, randomly drawn from our training data and scaled to the same S/N as the data in \cref{fig:lens-gallery}, i.e., our results data. The first sample is restricted to axis ratios $q>0.85$, and the second sample to $q<0.65$. In both we restrict the source position as a fraction of the Einstein radius to $\beta/\erad<0.1$. In this way the first sample consists only of `rings' and the second only of `quads'. We pass images of these systems through the model with no multipoles in training, $\modelone$.

\Cref{fig:multipole-orders} plots the results of this test. In the ring systems, it is more difficult for multipole perturbations to cause a false detection than in the other two samples. The high symmetry of the ring means it is difficult to disturb only one part of the lensed images without disturbing other parts, a signal which is not consistent with substructure. At larger multipole amplitudes, the false positive rate actually starts to decrease for the rings as they become completely disrupted by the perturbations. This point is important because it shows that our detector models do not simply classify as substructure anything which does not look like the negative case in the training data.

Comparing now the realistic sample and the quads, the $m=3$ mode behaves similarly in both cases. This is because $m=3$ perturbations can more easily disrupt individual images in a larger number of lensing configurations. Changing the configuration does not make a false positive more likely. However, in the $m=4$ mode the difference between the mixed and quad lenses is much larger, indicating that order four perturbations have a much stronger dependence on configuration. In a realistic sample of mixed rings, quads, and other configurations therefore, order three perturbations will always cause more false positives than order four.

\subsection{Comparison with classical method}
\label{sec:classical-comparison}
Sensitivity maps for mock HST lenses have also been produced with the `classical', i.e., non-machine learning method, in \citet{Despali2022}, which we refer to here as \citetalias{Despali2022}. We expect that, in general, the machine learning approach should produce more conservative sensitivity maps. This is because its prior for modelling an observation includes a much larger set of configurations and sources than in the classical method, where the prior is specific to the system being modelled. The machine learning method's prior is the entire set of training data, and some examples there may provide good fits to the data which the classical method would not allow for. We cannot make a direct comparison between this work and \citetalias{Despali2022} because of differences in data used, detection thresholds, and substructure concentration. However, we can show agreement with the main conclusions of that work.

\citetalias{Despali2022} shows primarily that the angular resolution sets the ultimate depth of the sensitivity. Higher S/N then increases the sensitivity everywhere up to this limit. We find a similar result when comparing this work to sensitivity maps for Euclid data in \citetalias{ORiordan2023}. We found a minimum sensitivity of $\Mmax=10^{8.8}\Msun$ at $3\sigma$ for Euclid VIS observations. In the HST-like observations used here, the minimum sensitivity at $3\sigma$ is $\Mmax=10^{7.8}\Msun$. These two minima represent the best-case scenario for each instrument in terms of S/N, lensing configuration, source complexity etc. The improvement in sensitivity of one order of magnitude can therefore be attributed to the increase in resolution, going from a pixel scale of 0.1 arcsec and PSF FWHM of 0.16 arcsec in Euclid VIS, to 0.04 arcsec and 0.07 arcsec in HST. This is in agreement with the trend in fig. 6 of \citetalias{Despali2022}.

In terms of S/N, a comparison is more difficult. \citetalias{Despali2022} analysed increasing S/N in the same systems, whereas we have a variety of systems with different S/N. We do find a weak trend with S/N and sensitive area, but it is confounded by other factors, e.g. configuration, redshifts, source structure, which would have to be controlled for to make a proper comparison. The effect of increasing S/N is mostly to increase the sensitive area, exactly the opposite of the effect due to including multipoles. This reinforces the need for observations of better angular resolution, rather than higher S/N.

The structure of the sensitivity maps produced with the two methods are remarkably similar. Here, we identify generic features of the sensitivity maps which are common to the methods, and can be seen in comparing this work's \cref{fig:sensitivity-maps} to fig. 2 of \citetalias{Despali2022}.
\begin{enumerate}
	\item For any configuration, sensitivity is strongly suppressed in the centre of the lens, where any small change in mass can be replicated by a change in the total mass, or the Einstein radius, of the lensing galaxy.
	\item In a non-symmetrical arc/counter-image system, i.e. when the source is near to, and inside, the fold of the caustic, sensitivity is weaker in counter-images compared to that on the main arc. This is clear in e.g. systems 37 and 40 here, or in the mocks labelled M4 and M9 in \citetalias{Despali2022}.
	\item The opposite of the previous statement is true in symmetric arc/counter-image systems, i.e. those with a source near to, and inside, the cusp of the caustic. In these sytems, sensitivity on the counter image is the same as that on the main arc. An example in this work is system 75, or M1 and M7 in \citetalias{Despali2022}.
	\item In the latter scenario, i.e., a cusp system, sensitivity on the arc peaks both at the locations of the images, but also at the points between the two outer images and the central one. These peaks of sensitivity are coincident with the critical curve in a purely elliptical system.
\end{enumerate}

\subsection{Comparison with other work}

Two dark matter substructures have so far been detected with gravitational imaging. In SDSSJ0946+1006 \citep[][also called the `double ring']{Vegetti2010} and in JVAS B1938+666 \citep{Vegetti2012}. The lens modelling in these cases did not include multipole perturbations so it would be natural to consider these detections in the context of this work. In gravitational imaging, a detection is made following strict criteria. We reprint those from \citet{Vegetti2014} here.
\begin{enumerate}
	\item The potential corrections must show a localised overdensity which improves the fit to the data and does not depend on regularisation conditions. 
	\item By parametric fitting, a substructure must be consistent with both the mass and the position of the found overdensity.
	\item The parametric model including the substructure must be favoured with an increase in Bayesian log-evidence of 50. This very high detection threshold is found to minimise false positive detections from parametric fitting \citep{Ritondale2019}.
\end{enumerate}
In this work we have, in effect, only considered a detection as that which satisfies (iii), although likely with a lower detection significance. In practice, a gravitational imaging study of a system from this work would not satisfy a strict application of (i) or (ii). Potential corrections are free-form and can in theory take on any part of the lensing potential which has not been included in the parametric model, including multipole perturbations. In both systems, the corrected density maps do not show evidence of multipole perturbations.

It should also be noted that both detections are located in positions which this work has shown to be particularly robust to the multipole degeneracy. The favoured substructure position is at the end of an arc in J0946, and at the centre of an arc in B1938. \Cref{fig:sensitivity-maps}, as well as \cref{fig:sensitivity-statistics} indicate that the depth of sensitivity in these locations would not change were multipoles included in the modelling.

Methods which do not use gravitational imaging to detect individual objects, i.e., the parametric methods we described in \cref{sec:introduction}, are more vulnerable to the degeneracy if it is not allowed for in the parametric model. Indeed, \citet{Nightingale2022} showed that four substructure detections in a sample of 54 HST strong lenses could be explained by missing angular complexity in the lens mass model. It is not possible within the scope of this work to say how the degeneracy with multipoles might effect an analysis based on flux ratio anomalies.

\subsection{Recommendations for strong lensing studies}
To maintain reliability when using strong lensing to infer the nature of dark matter, non-elliptical angular structure must be included in the lens model. When doing this, a prior must be chosen for the multipole orders and amplitudes. Many authors have adopted a normal prior with mean zero and standard deviation 1 per cent \citep{Gilman2023,Powell2022,Ballard2023}. This is an appropriate choice for two reasons.

First, it allows for the complexity observed in elliptical galaxy isophotes, which we briefly reviewed in \cref{sec:introduction}. Elliptical galaxies can show stronger multipole amplitudes, but not often in the region where strong lensing is sensitive. The multipole parameters are also not seen to evolve over the range of redshifts occupied by strong lenses, i.e., $z<1$, a finding corroborated by simulations. For lenses at higher redshifts which are likely to be found in upcoming surveys, this assumption may need to be reconsidered. This prior may also be problematic for lenses in cluster environments, where deviations from pure ellipticity may differ compared to isolated galaxies.

Second, and perhaps more importantly, it covers the precise region where mutlipole perturbations and substructure are degenerate. Consider again \cref{fig:multipole-orders}. In all three samples, the false positive rate only increases up to a point. In the mixed and quad samples, the FPR increases until an amplitude $\mstrength\sim0.03$, where it then plateaus. In the ring sample, it increases only until $\mstrength\sim0.02$. After these points, it can be assumed that multipole perturbations of increasing strength do not become more degenerate with substructure due to their larger size, and in full rings, they begin to look like something else entirely. In terms of adequately including the degeneracy with substructure then, amplitudes up to only a few per cent need be allowed. If a lens model was found to favour a perturbation of a higher amplitude, it is unlikely to be the result of a missing substructure.

	\section{Conclusions}
\label{sec:conclusions}

Strong gravitational lensing has emerged as one of the most promising tools for probing the nature of dark matter. A number of methods rely on strong lensing data to infer the properties of an underlying dark matter model, either directly or via the (non-)detection of dark substructures. In all cases, the lens galaxy mass model is a significant source of systematic uncertainty.

In this paper we focussed on angular structure beyond the elliptical power-law, parametrised with multipole perturbations up to and including order four. We used a sample of 100 realistic HST mock strong lens observations, similar in S/N and configuration to those used in previous substructure studies. Using a machine learning method developed previously we were able to produce sensitivity maps for these mock observations and test the effect of angular complexity in the mass model on substructure detection. We also compared our method to the classical sensitivity map method of \citet{Despali2022}, finding agreement where a comparison was possible. The main findings can be summarised as follows.
\begin{enumerate}
	\item Multipole perturbations to the lens galaxy mass model can be mistaken for substructure at a significant rate. The strength and order of  the perturbations which cause these false positives are similar to those measured in the isophotes of elliptical galaxies.
	\item The ability of such a perturbation to cause a false positive depends on the lensing configuration, with high-symmetry configurations like rings having the weakest degeneracy. Low-symmetry configurations such as quads have the strongest degeneracy.
	\item This relationship with configuration means that $m=3$ perturbations have a stronger effect than $m=4$. This is because $m=3$ can more easily cause a localised change to the images in average lensing configurations. However, both orders 3 and 4 have a strong effect in general.
	\item Sensitivity maps show that allowing multipoles in the lens mass model mainly reduces the area in the observation where a substructure could have been detected. When one per cent perturbations are allowed, the sensitive area drops by a factor of three.
	\item The depth of sensitivity, i.e., the minimum mass that could be detected, on the lensed images changes very little when multipoles are included in the model. In all models, with and without multipoles, we find a detection limit of $\Mmax\sim10^{8.2}\Msun$ at $5\sigma$ in HST data.
	\item The number of detections expected in HST quality data drops by $\sim 80$ per cent, assuming CDM, when multipoles of one per cent are allowed for in the model. This large drop is due to the loss of sensitive area in which only high-mass subhaloes were previously detectable. Although these subhaloes are less numerous in CDM, the large area on the sky in which they were previously detectable meant they contributed a large fraction of the total available detectable objects.
\end{enumerate}

We also discussed the implications of this work for previous substructure detections. We concluded that, for reasons relating to point (v), and others, they should be unaffected.

The points above have mixed consequences for the ability of strong lensing to uncover the nature of dark matter. A smaller number of detectable objects means poorer constraints overall on the total amount of substructure in dark matter haloes, i.e. on the normalisation of the SHMF, $\fsub$. Larger samples of lenses will be needed to constrain this parameter, compared to modelling without angular complexity. For low-mass subhaloes, the number of detections, albeit small to begin with, stays largely unchanged when multipole perturbations are included in the model. When attempting to discriminate between dark matter models, i.e., constraining $\Mhm$ or a particle mass, these low-mass detections are the most informative. However, in a regime where only relatively large subhaloes are detectable, e.g., in the optical data examined here, $\Mhm$ is degenerate with $\fsub$. The normalisation $\fsub$ may also vary across lens galaxies.
	
To alleviate these issues, a small sample of highly sensitive, high angular resolution observations could be combined with a large sample of less sensitive, low angular resolution observations. The sensitivity of the former provides stronger constraints on $\Mhm$, while the large numbers of the latter constrain $\fsub$. To achieve useful constraints on both parameters, such a complimentary strategy is required.

	\section*{Acknowledgements}
	
	CO'R thanks Giulia Despali and Simon White for helpful comments on this work. SV thanks the Max Planck Society for support through a Max Planck Lise Meitner Group. SV acknowledges funding from the European Research Council (ERC) under the European Union's Horizon 2020 research and innovation programme (LEDA: grant agreement No 758853).
	
	\section*{Data Availability}
	The data used in this paper are available from the corresponding author on request.

	\bibliographystyle{mnras}
	\bibliography{bibliography}
	
	\bsp	
	\label{lastpage}
	
\end{document}